# *New Pressure-Induced Polymorphic Transitions of Anhydrous Magnesium Sulfate*

A. Benmakhlouf[1], D. Errandonea[2], M. Bouchenafa[3], S. Maabed[1], A. Bouhemadou[4], A. Bentabet[5]

[1]*Laboratoire de Physique des Matériaux, Université Amar Telidji, BP 37G, Laghouat 03000, Algeria*

[2]*Departamento de Física Aplicada-ICMUV, MALTA Consolider Team, Universitat de Valencia, Edificio de Investigación, c/Dr. Moliner 50, Burjassot, 46100 Valencia, Spain*

[3]*Laboratoire de Physico-Chimie des matériaux, Université Amar Telidji ; BP 37G, Laghouat 03000, Algérie*

[4]*Laboratory for Developing New Materials and their Characterization, Department of Physics, Faculty of Science, University of Setif 1, 19000 Setif, Algeria*

[5]*Laboratoire de Recherche : Caractérisation et Valorisation des Ressources Naturelles, Université de Bordj Bou Arreridj, 34000, Algeria*

**Abstract:** The effects of pressure on the crystal structure of the three known polymorphs of magnesium sulfate (α-$MgSO_4$, β-$MgSO_4$, and γ-$MgSO_4$) have been theoretically study by means of density-functional theory calculations up to 45 GPa. We determined that at ambient conditions γ-$MgSO_4$ is an unstable polymorph, which decompose into MgO + $SO_3$, and that the response of the other two polymorphs to hydrostatic pressure is non-isotropic. Additionally we found that at all pressures β-$MgSO_4$ has a largest enthalpy than α-$MgSO_4$. This indicates that β-$MgSO_4$ is thermodynamically unstable versus α-$MgSO_4$ and predicts the occurrence of a β−α phase transition under moderate compression. Our calculations also predict the existence under pressure of additional phase transitions to two new polymorphs of $MgSO_4$, which we named as δ-$MgSO_4$ and ε-$MgSO_4$. The α−δ transition is predicted to occur at 17.5 GPa, and the δ−ε transition at 35 GPa, pressures that nowadays can be experimentally easily achieved. All the predicted structural transformations are characterized as first-order transitions. This suggests that they can be non-reversible, and therefore the new polymorphs could be recovered as metastable polymorphs at ambient conditions. The crystal structure of the two new polymorphs is reported. In them, the coordination number of sulfur is four as in the previously known polymorphs, but the coordination number of magnesium is eight instead of six. In the article we will report the axial and bond compressibility for the four polymorphs of $MgSO_4$. The pressure-volume equation of state of each phase is also given, which is described by a third-order Birch-Murnaghan equation. The values obtained for the bulk modulus are 62 GPa, 57 GPa, 102 GPa, and 119 GPa for α-$MgSO_4$, β-$MgSO_4$, δ-$MgSO_4$, and ε-$MgSO_4$, respectively. Finally, the electronic band structure of these four polymorphs of $MgSO_4$ has been calculated by the first time. The obtained results will be presented and discussed.

# 1 Introduction

The properties of inorganic materials under high pressure (HP) are of great fundamental and applied relevance. In particular, the effect of HP in oxides is a topic broadly studied in the present days. Among oxides, the HP study of sulfates has attracted considerable attention during the last decade.[1-5] Part of this interest comes from the fact that some sulfates have a crystal structure related to olivine, of the Earth's most common minerals, and therefore they can be used as test models to understand the high-pressure behavior of Earth´s mantle silicates. On the other hand, additional curiosity on the study of sulfates is related to their large number of applications which include catalyst agents, drilling fluids, food preparation, medical uses, pigments, and radiocontrast materials, among others.

Previous high-pressure studies on sulfates have been focused on $BaSO_4$ (the mineral barite),[1,2] $PbSO_4$ (the mineral anglesite),[3] $SrSO_4$ (the mineral celestite),[3] and $CaSO_4$ (the mineral anhydrite).[4,5] In all these compounds several phase transitions have been reported to take place 45 GPa. These transitions involve an important atomic rearrangement with changes in the coordination polyhedral units and the consequent effects on many physical and chemical properties.[1-5] Though there are several reports on the properties of magnesium sulfate, $MgSO_4$, they are mainly focused on its properties at ambient pressure and applications,[6-8] the crystal structure of the different polymorphs,[9] and the effect of temperature on it.[9] High-pressure studies on $MgSO_4$ have not carried out yet. Therefore, nothing is known on the possible existence of pressure-induced phase transitions, and the information on the compressibility of the different polymorphs is limited to the estimation of the bulk modulus from the thermal expansion using a Debbye approximation.[9] Part of the lack of HP studies is due to the fact that anhydrous $MgSO_4$ is strongly hygroscopic, which suppose experimental difficulties for the performance of such experiments.[10] Consequently, the performance of density-functional theory (DFT) calculations can be a good strategy to study the behavior of $MgSO_4$ under compression.

DFT has been used in the past to study the HP behavior of $BaSO_4$ and $CaSO_4$ showing that this method gives an accurate description of it.[1,4] It has been also used to carry out HP studies in other ternary oxides leading to predictions later confirmed by experiments. The list of oxides studied under HP by DFT calculations includes phosphates,[11] vanadates,[12] chromates,[13] tantalates,[14] molybdates,[15] and tungstates,[16] among others. The accuracy of the results obtained applying DFT to these compounds suggests that it could be useful to apply it to anhydrous $MgSO_4$. In our case, we employed this technique to theoretically study the HP behavior of the known polymorphs of $MgSO_4$. We explore their structural stability and their compressibility. Our results support the existence of phase transition to two novel polymorphs, which are isomorphic to the zircon[17] and scheelite[18] crystal structures. Finally, we also obtained information on the electronic band structure of the stable polymorphs of $MgSO_4$. The reported results are of interest for basic and applied research. They can be relevant for Planetary Sciences since $MgSO_4$ has been found in sulfate deposits in Mars[19] and in Europa one of the moon of Jupiter.[20]

**2 Computational Methods**

Total-energy calculations were performed at zero temperature (T = 0 K) as a function of pressure, up to 45 GPa, using DFT as implemented in the Cambridge Serial Total Energy Package (CASTEP).[21] The exchange-correlation energy was computed using the generalized-gradient approximation (GGA) with the Perdew-Burke-Ernzerhof parametrization developed for densely-packed solids (PBEsol).[22] The interaction between the valence electrons and the ion cores was properly described using the pseudo-potential plane-wave (PP-PW) method. The $Mg(2p^6, 3s^2)$, $O(2s^2, 2p^4)$, and $S(2p^6, 3s^2, 3p^4)$ electrons were considered explicitly as valence electrons and the remaining electrons and the nuclei were modelled using Vanderbilt ultra-soft pseudopotentials.[23] A kinetic energy cutoff of 500 eV and a 4x4x4 k-mesh were employed in the calculations. These parameters were selected after performing careful convergence tests. The optimized structural parameters were obtained using the Broyden-Fletcher-Goldfarb-Shanno (BFGS) minimization algorithm.[24] The crystal structure relaxation was carried out imposing four criteria for convergence: (i) the total-energy variation was smaller than $5.0 \times 10^{-6}$ eV/atom; (ii) the absolute value of force on any atom was less than 0.01 eV/Å; (iii) the stress was smaller than 0.02 GPa; and (iv) the atomic displacements were smaller than 0.0005 Å. From the computer simulations we obtained total energy (E) as a function of volume (V) and the corresponding pressure ($P = -\partial E/\partial V$) from which we determined the enthalpy (H). The thermodynamically stable phase at different pressures was obtained from the P−H curves of the analyzed candidate crystal structures. As candidate structure we assumed the polymorphs of $MgSO_4$ reported in the literature [9] together with the zircon and scheelite structures which are potential HP structures for $MgSO_4$ according with crystal-chemistry arguments.[25] This methodology has been previously applied successfully to the study of several ternary oxides under HP conditions.[26,27] Finally, the electronic band structure and density of states have been also calculated for the different polymorphs of $MgSO_4$. The obtained values for the band-gap energy ($E_g$) should be considered as a lower boundary for it due to the well-known tendency of GGA to slightly underestimate $E_g$.[28-30]

## 3 Results and discussion

### 3.1 Crystal structure and phase transitions

There are three polymorphs of $MgSO_4$ reported in the literature.[9] The ambient pressure stable polymorph ($\alpha$-$MgSO_4$) has an orthorhombic crystal structure with space group *Cmcm*.[31] There is also a high-temperature (HT) modification of $MgSO_4$ which is metastable at ambient conditions.[32] Its structure is also orthorhombic but it belongs to space group *Pnma*.[33] The crystal structure of both polymorphs is shown in **Fig. 1**. Both structures are formed by $MgO_6$ octahedral units and $SO_4$ tetrahedral units. An accurate description of them has been provided by Fortes *et al.*[9] There is additionally a third polymorph reported for $MgSO_4$ in the literature, $\gamma$-$MgSO_4$. This polymorph has been apparently detected at HT above 1273 K.[34] However, it is known that $MgSO_4$ decomposes at a lower temperature. This polymorph has been never confirmed and never found as a metastable phase at ambient conditions. It has been proposed that $\gamma$-$MgSO_4$ could have a crystal structure similar to that of $\gamma$-$CoSO_4$.[35] Considering this structure for $MgSO_4$ we have found that $\gamma$-$MgSO_4$ has a quite higher enthalpy than $\alpha$-$MgSO_4$ and $\beta$-$MgSO_4$. In addition, we found that the enthalpy of $\gamma$-$MgSO_4$ is larger than that of $MgO + SO_3$. Therefore, it is quite unlikely that $\gamma$-$MgSO_4$ could be observed at room temperature. Consequently, we have not included this polymorph in the high-pressure calculations.

**Fig. 2** shows the calculated pressure dependence of the enthalpy difference between the two candidate HP phases of $MgSO_4$ (zircon-type $\delta$-$MgSO_4$ and scheelite-type $\varepsilon$-$MgSO_4$) and the known ambient-pressure polymorphs ($\alpha$-$MgSO_4$ and $\beta$-$MgSO_4$). The stable $\alpha$-$MgSO_4$ is used as the enthalpy of reference in the figure. Based upon thermodynamics, the structure with the lowest enthalpy is assumed to be the stable structure at different pressure. The selection of the HP candidate structure has been made using crystal-chemistry arguments, like the diagram proposed by Bastide,[36] which has been successfully used to understand and predict HP phase transitions in many ternary oxides.[25] According to Bastide, pressure-induced transitions are expected to occur from the crystal structure of a given compound to that of a larger-cation-hosting compound. Therefore, the tetragonal zircon structure[17] (space group *I4₁/amd*) of ZrSiO₄ and the tetragonal scheelite structure[18] (space group *I4₁/amd*) of CaWO₄ were assumed as structural models for δ-MgSO₄ and ε-MgSO₄, respectively. **Fig. 2** shows α-MgSO₄ as being the most stable phase at zero and low pressure. At ambient pressure the enthalpy of β-MgSO₄ is approximately 0.1 eV larger than that of α-MgSO₄. Thus, β-MgSO₄ is observed as a metastable phase at ambient conditions due to the presence of kinetic barriers that prevents its transformation into β-MgSO₄. As pressure increases the enthalpy difference is enlarged, becoming β-MgSO₄ thermodynamically highly unstable. Therefore, it is quite possible that a compression of a few GPa would transform β-MgSO₄ into α-MgSO₄. Note that a similar situation has been reported for BiPO₄.[37] In this compound a hydrostatic compression of 3 GPa is enough to transform the metastable SbPO₄-type polymorphs into the stable monazite-type polymorph. In **Table 1** we present the calculated structural parameters of α-MgSO₄. In **Table 2** we present the calculated structural parameters of β-MgSO₄. The agreement with the crystal structure determined from neutron powder diffraction and single-crystal x-ray diffraction is very good not only for lattice parameters but also for atomic positions.[9] Differences are within 2% which is typical for DFT calculations.[38] Our results are also comparable with previous ambient-pressure DFT calculations carried out using the Vienna Ab-initio Simulations Package (VASP).[39]

As pressure increases, α-MgSO₄ becomes unstable against the two proposed HP polymorphs. At 17.5 GPa δ-MgSO₄ becomes the polymorph with the lowest enthalpy. Beyond 35 GPa ε-MgSO₄ becomes the polymorph with the lowest enthalpy. These results support the occurrence of two pressure-induced phase transitions in MgSO₄. The first one takes place at 17.5 GPa from α-MgSO₄ to δ-MgSO₄ and the second one occurs at 35 GPa from δ-MgSO₄ to ε-MgSO₄. The structural information of both HP phases is summarized in **Tables 3 and 4.** Notice than the δ-MgSO₄ to ε-MgSO₄ transition is analogous to the zircon to scheelite transition that has been observed in many ternary oxides.[40] The crystal structure of the two new polymorphs is shown in **Fig. 2**. The successive phase transition involves a densification of MgSO$_4$ and an increase in the coordination number of Mg, which increases from six in α-MgSO$_4$ and β-MgSO$_4$ to eight in δ-MgSO$_4$ and ε-MgSO$_4$.

### 3.2 Compressibility and P-V equations of state

From our calculations we have obtained the pressure dependence of the unit-cell parameters of the different polymorphs of MgSO$_4$. The results for the pressure range of stability of each phase are shown in **Fig. 3**. In the case of β-MgSO$_4$ results are given up to 10 GPa since the transition from this polymorphs to α-MgSO$_4$ is expected to take place under hydrostatic compression at low pressures as explained previously. In the figure it can be seen that the compressibility of in α-MgSO$_4$ and β-MgSO$_4$ is anisotropic, being the *a*-axis the less compressible axis in both cases. This asymmetric behavior of the different crystallographic axes is also revealed in their different thermal expansion.[9] This observation is not surprising since in many ternary oxides the compressibility and the thermal expansion are closely related with each other.[41]

From the results shown in **Fig. 3**, we determined the axial compressibilities at ambient pressure in the different crystallographic directions, $k_x = \frac{-1}{x}\frac{\partial x}{\partial P}$, where $x$ = *a*, *b*, or *c*. The obtained values are summarized in **Table 5**. In the table we also present the ratio between the different compressibilities to illustrate the anisotropic behavior of the different polymorphs. These ratios show that clearly the *a*-axis is much less compressible than the other axes. Interestingly, as a consequence of the anisotropic compression α-MgSO$_4$ becomes gradually more symmetric under compression, but the asymmetry of β-MgSO$_4$ is enhanced under pressure. From the obtained values for the axial compressibilities the bulk modulus (B$_0$) can be determined as $B_0 = \frac{1}{k_a+k_b+k_c}$. We have obtained $B_0$ = 70 GPa for α-MgSO$_4$ and $B_0$ = 58 GPa for β-MgSO$_4$. Thus the second polymorphs is more compressible than the first one, which is consistent with the fact that α-MgSO$_4$ is more compact (has a smaller volume) than β-MgSO$_4$.

For the two new polymorphs, we found that they are less compressible than the ambient-pressure polymorphs. This can be seen in **Fig. 4** and **Table 5**, where we show the calculated axial compressibilities for δ-MgSO$_4$ at 18 GPa and for ε-MgSO$_4$ at 36 GPa. We also determined that the contraction of δ-MgSO$_4$ and ε-MgSO$_4$ is not isotropic. However, the differences in their axial compressibilities are not as large as in α-MgSO$_4$ and β-MgSO$_4$. This is illustrated in **Fig. 4** and also in **Table 5**. In the case of zircon-type δ-MgSO$_4$ the most compressible axis is *a*. In the case of scheelite-type ε-MgSO$_4$ the most compressible axis is *c*. Therefore in δ-MgSO$_4$ *c/a* increases under compression as occurs in most zircon-structured oxides[17] and in ε-MgSO$_4$ *c/a* decreases under compression as occurs in most scheelite-structured oxides[18]. Again, from the axial compressibilities we estimated the bulk modulus for δ-MgSO$_4$ at 18 GPa, whose value is 193 GPa, and the bulk modulus of ε-MgSO$_4$ at 36 GPa, whose value is 266 GPa.

In **Fig. 3** we also show the pressure dependence of the unit-cell volume of the different polymorphs. There is can be seen that the polymorphs becomes less compressible following the sequence β-MgSO$_4$ < α-MgSO$_4$ < δ-MgSO$_4$ < ε-MgSO$_4$. Thus, as expected, there is a direct relation between the volume of the polymorph and its compressibility. The largest the unit-cell volume the most compressible the polymorph is. This is exactly what is expected since for polymorphs of a same material (i.e. with similar bonding properties) the bulk modulus would scale inversely with the unit-cell volume.[42] In the figure, it can also be seen that the different phase transition involves a volume collapse of the crystal structure. The relative volume contraction when going from β-MgSO$_4$ to α-MgSO$_4$ is 3.4 %, the relative volume contraction when going from α-MgSO$_4$ to δ-MgSO$_4$ is 4 %, and the relative volume contraction when going from β-MgSO$_4$ to α-MgSO$_4$ is 5 %. Such volume changes are consistent with first-order phase transitions.

From the results shown in **Fig. 3** we have determined the room-temperature P-V equation of state (EOS) for the different phases of MgSO$_4$. Using a normalized stress vs finite strain plot[43] we have determined that the pressure dependence of the volume can be well described by a third-order Birch-Murnaghan EOS[44] in the four phases. The obtained unit-cell volume at ambient pressure (V$_0$), bulk modulus (B$_0$), and its pressure derivative (B$_0$') are given in **Table 6**. The implied values of the second pressure derivative of the bulk modulus (B$_0$'') in a third-order truncation of the Birch-Murnaghan EOS[45] are also given. According with these results the bulk moduli α-MgSO$_4$ and β-MgSO$_4$ are comparable with those of BaSO$_4$ (58 – 63 GPa)[1] and CaSO$_4$ (64 GPa).[4] By comparing the results of our calculations with the estimations made using a Debbye model,[9] we conclude that this model overestimate by more than 30 % the bulk modulus of α-MgSO$_4$ and β-MgSO$_4$.

It has been shown than for many ternary oxides the bulk compressibility can be expressed by means of cation oxide polyhedral compressibilities.[46] In the case of phosphates,[12] vanadates,[13] and many other oxides[25] the small tetrahedral units (e.g. PO$_4$ or VO$_4$) are highly uncompressible and therefore the other cation oxide polyhedron (e.g. YO$_8$ in YPO$_4$ and YVO$_4$) is the one that dominates the bulk compressibility.[47,48] However, in the case of MgSO$_4$, CaSO$_4$, and BaSO$_4$ apparently the compressibility of the small SO$_4$ tetrahedral units cannot be neglected. This conclusion is supported by the fact that the three alkaline-earth sulfates mentioned above have a similar bulk modulus in spite that the compressibility of Mg-O, Ca-O, and Ba-O bonds is very different; as can be seen by the different bulk modulus of MgO (150 GPa),[49] CaO (115 GPa),[50] and BaO (75 GPa).[51] Additional evidence supporting this conclusion will be presented in the next section when reporting bond compressibilites. Notice that the compressible S-O bond makes the sulfates to have a bulk modulus considerable similar to chromates[13] and smaller than phosphates,[11] vanadates,[12] tantalates,[14] molybdates,[15] and tungstates.[16] Notice also that among the ternary oxides studied under compression only perchlorates[52] are more compressible than sulfates.

Regarding the HP phases of $MgSO_4$, it **Table 6** it can be seen that the bulk modulus of both δ-$MgSO_4$ and ε-$MgSO_4$ nearly doubles the value of the same parameter in the ambient-pressure polymorphs. Such a large increase is consistent with the changes observed when comparing the low-pressure and high-pressure phases of $BaSO_4$ and $CaSO_4$.[1,53] The reduction of the compressibility of $MgSO_4$ in the HP phases is also consistent with the volume collapse associated to each transition. To conclude, we would like to state that from the calculated EOS it can be estimated that the bulk modulus of δ-$MgSO_4$ at 18 GPa is 195 GPa, and that the bulk modulus of ε-$MgSO_4$ at 36 GPa is 274 GPa. Both values agree with those determined from the axial compressibilities.

### 3.3 Bond compressibility

The analysis of the effect of pressure in bond distances has been proved to be quite useful to understand the macroscopic properties of ternary oxides like $MgSO_4$.[54] In this section we will analyze the pressure dependence of the different bond distances of the four polyhedral of $MgSO_4$ and relate it with their compressibility. **Table 7** shows the calculated bond distances at selected pressure for each polymorph. The table also gives the coordination number for each cation, the distortion index for the bond length (as defined by VESTA),[55] and the average bond distance. **Fig. 4** shows the calculated pressure dependence of the bond distances. There it can be seen that in $\alpha$-MgSO$_4$ the two different S-O bonds changes with pressure in a similar way. As a consequence of it the shape on the SO$_4$ tetrahedron is basically not modified by pressure, changing only its volume. We found that the pressure dependence of the tetrahedron volume is consistent with a polyhedral bulk modulus of 70 GPa. In the case of the Mg-O bonds we found that the four long bonds are more compressible than the two short bonds. As a result of it, the MgO$_6$ octahedron becomes more regular under compression. Regarding the volume change of the octahedron, we found that its relative change is comparable with the relative change of the SO$_4$ tetrahedron. In particular, we determined that the polyhedral bulk modulus of the MgO$_6$ is 60 GPa. The similar bulk modulus of both kinds of polyhedral units is a probe that the SO$_4$ tetrahedral units are not rigid and play a role in the bulk compressibility.

In the case of $\beta$-MgSO$_4$ the behavior of bond distances is similar than in $\alpha$-MgSO$_4$. As can be seen in **Fig. 4**, the different S-O bonds behave in a similar way under compression. In contrast, for the Mg-O bonds we found that the two pairs of long bonds are more compressible than the short bonds. Thus, as happens in $\alpha$-MgSO$_4$, under HP the shape of the SO$_4$ tetrahedron is basically not modified (only the volume is reduced) but the shape of the MgO$_6$ octahedron is modified becoming it more symmetric as pressure increases. On the other hand, we found that both Mg-O and S-O bonds are slightly more compressible in $\beta$-MgSO$_4$ than in $\alpha$-MgSO$_4$. This can be clearly perceived in **Fig. 4**. This is consistent with the fact that $\beta$-MgSO$_4$ has a smaller bulk modulus than $\alpha$-MgSO$_4$ (see **Table 6**).

Comparing in **Fig. 1** $\alpha$-MgSO$_4$ and $\beta$-MgSO$_4$ with the HP polymorphs of MgSO$_4$, it can be seen than the phase transitions involves an important structural rearrangement. In the first place, in $\delta$-MgSO$_4$ the SO$_4$ tetrahedron becomes regular, with an S-O distance that is shorter than the average S-O distances in the low-pressure phases. In addition, the compressibility of this bond is reduced as can be seen in **Fig. 4**. Regarding Mg, it can be seen that its coordination number increases from six to eight. In particular an $MgO_8$ dodecahedron is formed. On it there are four short bond with identical length and four long bonds with identical length, being the average interatomic distances enlarged in comparison with that of the $MgO_6$ octahedron of the low-pressure phases. This a consequence of the accommodation of two extra oxygens within the coordination sphere of Mg. On the other hand, the compressibility of the Mg-O bonds is also reduced in $\delta$-$MgSO_4$. This and the reduction of the compressibility of the S-O bonds explain the large increase of the bulk modulus in $\delta$-$MgSO_4$.

Finally, the structural changes observed at the $\delta$-$MgSO_4$ to $\varepsilon$-$MgSO_4$ transition are those typical of a zircon-scheelite transition. The structural relation between both crystal structures has been described by Nyman *et al.*[56] In particular, the schellite structure ($\varepsilon$-$MgSO_4$) implies an increase of the packing efficiency of the crystal structure. As can be seen in **Fig. 4**, the phase transition involves a small enlargement of the S-O distance, a reduction of the long Mg-O distance, and an enlargement of the short Mg-O distance. As a consequence of it, the $MgO_8$ dodecahedron becomes more symmetric in $\varepsilon$-$MgSO_4$ than it is in $\delta$-$MgSO_4$. This is illustrated by the decrease of the distortion index which is approximately the double in $\delta$-$MgSO_4$ than in $\varepsilon$-$MgSO_4$. Regarding bond compressibilities, **Fig. 4** shows that the compressibility of S-O and Mg-O are reduced in $\varepsilon$-$MgSO_4$ which justifies the 15 % increase of the bulk modulus associated to the phase transition.

### 3.4 Electronic properties

Optical properties of solids are a major topic, both in basic research as well as for industrial applications. In the case of anhydrous $MgSO_4$, the band-gap has not been experimentally determined yet. On the other hand, band-structure calculations have been carried out only for α-MgSO$_4$.[39,57] The calculations carried out using the program CRYSTAL03 gave an energy band gap (E$_g$) of 7.4 eV.[57] The calculations performed using VASP leads to E$_g$ = 6.0 eV,[39] which is considerable smaller. In our case, we have calculated the band structure and electronic density of states (DOS) for the four polymorphs of MgSO$_4$ here studied under compression. The results are shown in **Figs. 5 to 8**. For α-MgSO$_4$ we determined that it is a wide band-gap semiconductor with a direct band gap at the Γ point of the Brillouin zone (see **Fig. 5**). The determined value of E$_g$ (5.457 eV) is more similar to the value determined using VASP[39] than to the value obtained using CRYSTAL03.[57] Regarding other important features of the band structure, we would like to remark that the dispersion of the valence bands is relatively small, with comparable dispersions along different directions. We also determined that the upper half of the valence band consists of purely non-bonding O 2p states. On the other hand, the bottom of the conduction band is dominated by Mg 2p and 3s and S 3s electrons, with a small contribution of O 2p states.

In **Fig. 6** the results obtained for β-MgSO$_4$ are reported. In the figure it can be seen that both the band structure and the DOS are qualitatively similar to those of β-MgSO$_4$. This observation is not surprising giving the structural similarities between both polymorphs. However, β-MgSO$_4$ is found to be an indirect band-gap material, with the top of the valence band at the X point of the Brillouin zone and the bottom of the conduction band at the Γ point. The determined value of E$_g$ is 5.495 eV, a value only slightly larger than the energy band gap of α-MgSO$_4$. These values indicate that the two ambient-pressure polymorphs of MgSO$_4$ have a band gap considerable larger than related chromates, molybdates, tungstates, and selenates.[58] This is a consequence of the circumstance that classical field splitting in the SO$_4^{2-}$ ion is larger than in other ions like SeO$_4^{2-}$ or WO$_4^{2-}$.[58] On the other hand, the electronic band gaps of α-MgSO$_4$ and β-MgSO$_4$ are comparable with those of phosphates; e.g LaPO$_4$.[59]

Regarding δ-MgSO$_4$ and ε-MgSO$_4$, we found that both HP phases are direct band gap materials. Their band structures are qualitative similar to those of other zircon-structure (for δ-MgSO$_4$)[60] and scheelite structured (for ε-MgSO$_4$) oxides.[29] Both polymorphs are direct band-gap materials (see **Fig. 7 and 8**) with E$_g$ = 6.603 eV and 6.888 eV for δ-MgSO$_4$ at 18 GPa and ε-MgSO$_4$ at 36 GPa, respectively. In both polymorphs the band gap is at the Γ point of the Brillouin zone. With respect to the composition of the valence and conduction bands, in **Figs. 7 and 8** it can be seen that as in the low-pressure polymorphs, both in δ-MgSO$_4$ and ε-MgSO$_4$ the top of the valence band is dominated by O 2p states. On the other hand, in the conduction band there is again a contribution of Mg, S, and O orbitals.

We have explored the pressure dependence of E$_g$ for the different polymorphs of MgSO$_4$. The results are shown in **Fig. 9**. We found that in the four polymorphs the band gap opens under compression. This is a consequence of the increase of the repulsion between bonding and anti-bonding states, which makes the conduction band to moves faster towards higher energies than the valence band. For α-MgSO$_4$ and β-MgSO$_4$ at ambient pressure we determined the pressure coefficients for the band-gap energy to be dE$_g$/dP = 54 meV/GPa and 58 meV/GPa, respectively. Additionally, at the α-to-δ transition we found that the structural changes cause and opening of the band gap. In particular, for α-MgSO$_4$ the value of E$_g$ at 18 GPa is 6.427 eV, while for δ-MgSO$_4$ at the same pressure E$_g$ = 6.603 eV. This increase of E$_g$ by approximately 0.2 eV should be clearly detected by experiments. At 18 GPa, the pressure coefficient of E$_g$ for δ-MgSO$_4$ is 47 meV/GPa. The decrease of this coefficient is consistent with the compressibility decrease associated to the transition which was discussed in **section 3.3**. Finally, at the δ-to-ε transition our calculations predict a collapse of the band gap. At 36 GPa E$_g$ is 7.310 eV for δ-MgSO$_4$ and 6.888 eV for ε-MgSO$_4$. This large band-gap collapse (ΔE$_g$ = 0.42 eV) is typical of the zircon-scheelite transitions[60], like the δ-to-ε transition.

Regarding the pressure coefficient of $E_g$, at 36 GPa for ε-MgSO$_4$ we determined $dE_g/dP$ = 37 meV/GPa. Consequently, among the four polymorphs, the one that has the smallest $dE_g/dP$ is ε-MgSO$_4$. This is a result of the fact that ε-MgSO$_4$ is the least compressible polymorph.

**4 Concluding Remarks**

We performed a first principle computational analysis of the high-pressure properties of anhydrous MgSO$_4$ up to 45 GPa. In particular, we exhaustively and systematically study of the effect of HP in the crystal and electronic structures of MgSO$_4$. At pressures that can be routinely achieved now-a-days we predict the existence of two structural phase transitions to two previously unknown polymorphs. These polymorphs are isomorphic with zircon and scheelite structures. The crystal structure of the different polymorphs is reported. Calculations also allowed us to obtain information on the axial, bond, and bulk compressibility of different polymorphs. The P-V EOS for the four studied polymorphs has been determined. We also calculated the electronic band structure and density of states for the four studied polymorphs. We found that all of them are wide band-gap semiconductors. The pressure dependence of $E_g$ has been also obtained. The reported results are compared with previous studies in related oxides. We hope our finding with trigger further studies on the HP behavior of MgSO$_4$.

**Acknowledgements**

This study was partially supported by the Spanish Ministerio de Economıa y Competitividad (MINECO) under Grants No. MAT2016-75586-C4-1-P and No.MAT2015-71070-REDC (MALTA Consolider). Part of the calculations reported in this work were performed in the Laboratory de Caractérisation et Valorisation des Ressources Naturelles at Bordj Bou

Arreridj university. A. Benmakhlouf and A. Bentabet would like to thank the help of various members of this laboratory.## References

(1) Santamaria-Perez, D; Gracia, L.; Garbarino, G.; Beltran, A.; Chulia-Jordan, R.; Gomis, O.; Errandonea, D.; Ferrer-Roca, Ch.; Martinez-Gracia, D.; Segura A. *Phys. Rev. B* **2011**, *84*, 054102.

(2) Crichton, W.A.; Merlini, M.; Hanfland, M.; Müller, H. *Amer. Mineral.* **2001**, *96,* 364.

(3) Lee, P.L.; Huang, E.; Yu, S.C. Yen-Hua Chen *World Journal of Condensed Matter Physics.* **2013**, *3,* 28.

(4) Gracia, L.; Beltrán, A.; Errandonea, D.; Andrés, J. *Inorg. Chem.* **2012**, *51*, 1751.

(5) Fujii, T.; Ohfuji, H.; Inoue, T. *Phys. Chem. Miner.* **2016**, *43*, 353.

(6) Ashalu, K.C.; Rao, J.N. *J. Chem. Pharm. Res.* **2013**, *5,* 44.

(7) Iype, E.; Nedea, S.V.; Rindt, C.M.; van Steenhoven, A.A., Zondag, H.A. Jansen, P.J. *J. Phys. Che. C* **2012**, *116*, 18584.

(8) Schidema, M.N.; Taskinen, P. *Ind. Eng. Chem. Res.* **2011**, *50*, 9550.

(9) Fortes, A.D.; Wood, I.G.; Vocadlo, L.; Brand, H.E.A. Knight K.S. *J. Appl. Cryst..* **2007**, *40*, 761.

(10) Errandonea, D. *Crys. Res. Tech.* **2015**, *50*, 729.

(11) Lopez-Moreno, S.; Errandonea, D. *Phys. Rev. B* **2012**, *86*, 104112.

(12) Garg, A.B.; Errandonea, D.; Rodríguez-Hernández, P.; López-Moreno, S.; Muñoz, A.; Popescu, C. *J. Phys.: Condens. Matter* **2014**, *26*, 265402.

(13) Gleissner, J.; Errandonea, D.; Segura, A.; Pellicer-Porres, J.; Hakeem, M.A.; Proctor, J.E.; Raju, S.V.; Kumar, R.S.; Rodrıguez-Hernandez, P.; Muñoz, A.; Lopez-Moreno, S.; Bettinelli, M. *Phys. Rev. B* **2016**, *94*, 134108.
17

# Figure Captions

**Fig. 1** Crystal structure of α-MgSO$_4$ (a), β-MgSO$_4$ (b), δ-MgSO$_4$ (c), and ε-MgSO$_4$ (d). The coordination polyhedra of S are shown in yellow and the coordination polyhedra of Mg in orange. The oxygen atoms are shown as small red spheres. Note that δ-MgSO$_4$ and ε-MgSO$_4$ involves an increase of the coordination number of Mg from 6 to 8.

**Fig. 2.** Calculated enthalpy difference versus pressure for the different polymorphs. The enthalpy difference has been calculated relative to α-MgSO$_4$.

**Fig. 3.** Pressure dependence of the unit-cell parameters and volume for the different polymorphs of MgSO$_4$.

**Fig. 4.** Calculated bond distance versus pressure in different polymorphs of MgSO$_4$.

**Fig. 5.** (Top) Calculated band structure of α-MgSO$_4$ at ambient pressure. (Bottom) Calculated total density of states (TDOS) and partial densities of states for Mg, S, and O.

**Fig. 6.** (Top) Calculated band structure of β-MgSO$_4$ at ambient pressure. (Bottom) Calculated total density of states (TDOS) and partial densities of states for Mg, S, and O.

**Fig. 7.** (Top) Calculated band structure of δ-MgSO$_4$ at 18 GPa. (Bottom) Calculated total density of states (TDOS) and partial densities of states for Mg, S, and O.

**Fig. 8.** (Top) Calculated band structure of ε-MgSO$_4$ at 36 GPa. (Bottom) Calculated total density of states (TDOS) and partial densities of states for Mg, S, and O.

**Fig. 9.** Pressure dependence of the band-gap energy for the different polymorphs of MgSO$_4$.

**Table Captions**

**Table 1.** Calculated crystal structure of α-MgSO$_4$ at ambient pressure and 0 K (space group *Cmcm*).

**Table 2.** Calculated crystal structure of β-MgSO$_4$ at ambient pressure and 0 K (space group *Pnma*).

**Table 3.** Calculated crystal structure of δ-MgSO$_4$ at 18 GPa and 0 K (space group *I4$_1$/amd*).

**Table 4.** Calculated crystal structure of ε-MgSO$_4$ at 35 and 0 K (space group *I4$_1$/a*).

**Table 5.** Theoretical values for axial compressibilities of the different polymorphs of MgSO$_4$ along with their ratios. The reported values are calculated at ambient pressure for α-MgSO$_4$ and β-MgSO$_4$, at 18 GPa for δ-MgSO$_4$, and at 36 GPa for ε-MgSO$_4$.

**Table 6.** EOS determined for the different polymorphs of MgSO$_4$. The volume (V$_0$), bulk modulus (B$_0$), its pressure derivative (B$_0$'), and the implied value of the second pressure derivative (B$_0$'') are given.

**Table 7.** The typical bond lengths calculated for the different phases of MgSO$_4$. The pressure corresponding to each polymorph is indicated. In addition to bond distances we also include other polyhedral parameters like coordination number (CN), distortion index, and average bond distance.

**Table 1.** Calculated crystal structure of α-MgSO$_4$ at ambient pressure and 0 K (space group *Cmcm*).

| | Site | x | y | z |
|---|---|---|---|---|
| $a = 5.2048$ Å, $b = 7.9893$ Å, $c = 6.5813$ Å | | | | |
| Mg | 4a | 0 | 0 | 0 |
| S | 4c | 0 | 0.35017 | 0.25 |
| O$_1$ | 8f | 0 | 0.24934 | 0.06410 |
| O$_2$ | 8g | 0.26543 | 0.96083 | 0.25 |

**Table 2.** Calculated crystal structure of β-MgSO$_4$ at ambient pressure and 0 K (space group *Pnma*).

| | Site | x | y | z |
|---|---|---|---|---|
| $a = 8.6947$ Å, $b = 6.7807$ Å, $c = 4.8010$ Å | | | | |
| Mg | 4a | 0 | 0 | 0 |
| S | 4c | 0.32101 | 0.25 | 0.02428 |
| O$_1$ | 8d | 0.37235 | 0.06900 | 0.16489 |
| O$_2$ | 4c | 0.14765 | 0.25 | 0.03160 |
| O$_3$ | 4c | 0.37928 | 0.25 | 0.73129 |

**Table 3.** Calculated crystal structure of δ-MgSO$_4$ at 18 GPa and 0 K (space group *I4$_1$/amd*).

| | Site | x | y | z |
|---|---|---|---|---|
| $a = 6.2388$ Å, $c = 5.6207$ Å | | | | |
| Mg | 4a | 0 | 0.75 | 0.125 |
| S | 4b | 0 | 0.25 | 0.375 |
| O | 16h | 0 | 0.43552 | 0.21462 |

**Table 4.** Calculated crystal structure of ε-MgSO$_4$ at 36 and 0 K (space group *I4$_1$/a*).

| | Site | x | y | z |
|---|---|---|---|---|
| $a = 4.34700$ Å, $c = 10.16910$ Å | | | | |
| Mg | 4b | 0 | 0.25 | 0.625 |
| S | 4a | 0 | 0.25 | 0.125 |
| O$_1$ | 16f | 0.24963 | 0.11039 | 0.04837 |

**Table 5.** Theoretical values for axial compressibilities of the different polymorphs of MgSO$_4$ along with their ratios. The reported values are calculated at ambient pressure for α-MgSO$_4$ and β-MgSO$_4$, at 18 GPa for δ-MgSO$_4$, and at 36 GPa for ε-MgSO$_4$.

| Phase | $\kappa_a$ (10$^{-3}$ GPa$^{-1}$) | $\kappa_b$ (10$^{-3}$ GPa$^{-1}$) | $\kappa_c$ (10$^{-3}$ GPa$^{-1}$) | $\kappa_b/\kappa_a$ | $\kappa_c/\kappa_a$ | $\kappa_c/\kappa_b$ |
|---|---|---|---|---|---|---|
| α-MgSO$_4$ | 1.33 | 5.41 | 7.55 | 4.07 | 5.68 | 1.40 |
| β-MgSO$_4$ | 1.59 | 7.75 | 7.83 | 4.87 | 4.92 | 1.01 |
| δ-MgSO$_4$ | 2.03 | | 1.13 | | 0.56 | |
| ε-MgSO$_4$ | 0.88 | | 2.08 | | 2.36 | |

**Table 6.** EOS determined for the different polymorphs of MgSO$_4$. The volume (V$_0$), bulk modulus (B$_0$), its pressure derivative (B$_0$'), and the implied value of the second pressure derivative (B$_0$'') are given.

| | V$_0$ (Å$^3$) | B$_0$ (GPa) | B$_0$' (dimensionless) | B$_0$'' (GPa$^{-1}$) |
|---|---|---|---|---|
| α-MgSO$_4$ | 273.8 | 62 | 5.0 | -0.095 |
| β-MgSO$_4$ | 282.7 | 57 | 5.7 | -0.149 |
| δ-MgSO$_4$ | 248.3 | 103 | 5.1 | -0.060 |
| ε-MgSO$_4$ | 233.8 | 120 | 4.3 | -0.036 |

**Table 7.** The typical bond lengths calculated for the different phases of MgSO$_4$. The pressure corresponding to each polymorph is indicated. In addition to bond distances we also include other polyhedral parameters like polyhedral volume, coordination number (CN), distortion index, and average bond distance.

| α-MgSO$_4$ | β-MgSO$_4$ | δ-MgSO$_4$ | ε-MgSO$_4$ |
|---|---|---|---|
| Ambient pressure | Ambient pressure | 18 GPa | 36 GPa |
| S-O$_1$ = 1.50738 Å (x2) | S-O$_1$ = 1.47012 Å (x2) | S-O = 1.46705 Å (x4) | S-O = 1.46734 Å (x4) |
| S-O$_2$ = 1.46485 Å (x2) | S-O$_2$ = 1.50772 Å | | |
| | S-O$_2$ = 1.49510 Å | | |
| <S-O> = 1.4861 Å | <S-O> = 1.4858 Å | | |
| V$_{SO4}$ = 1.6826 Å$^3$ | V$_{SO4}$ = 1.6796 Å$^3$ | V$_{SO4}$ = 1.6101 Å$^3$ | V$_{SO4}$ = 1.6061 Å$^3$ |
| CN = 4 | CN = 4 | CN = 4 | CN = 4 |
| Dist. index = 0.01431 | Dist. index = 0.01053 | | |
| Mg-O$_1$ = 2.03623 Å (x2) | Mg-O$_1$ = 2.00977 Å (x2) | Mg-O = 2.02561 Å (x4) | Mg-O = 2.06060 Å (x4) |
| Mg-O$_2$ = 2.17108 Å (x4) | Mg-O$_2$ = 2.13183 Å (x2) | Mg-O = 2.23238 Å (x4) | Mg-O = 2.15895 Å (x4) |
| | Mg-O$_3$ = 2.28219 Å (x2) | | |
| <Mg-O> = 2.1261 Å | <Mg-O> = 2.1413 Å | <Mg-O> = 2.1290 Å | <Mg-O> = 2.1098 Å |
| V$_{MgO6}$ = 12.5610 Å$^3$ | V$_{MgO6}$ = 12.4413 Å$^3$ | V$_{MgO8}$ = 17.1023 Å$^3$ | V$_{MgO8}$ = 16.5722 Å$^3$ |
| CN = 6 | CN = 6 | CN = 8 | CN = 8 |
| Dist. index = 0.02819 | Dist. index = 0.04388 | Dist. index = 0.04856 | Dist. index = 0.02331 |

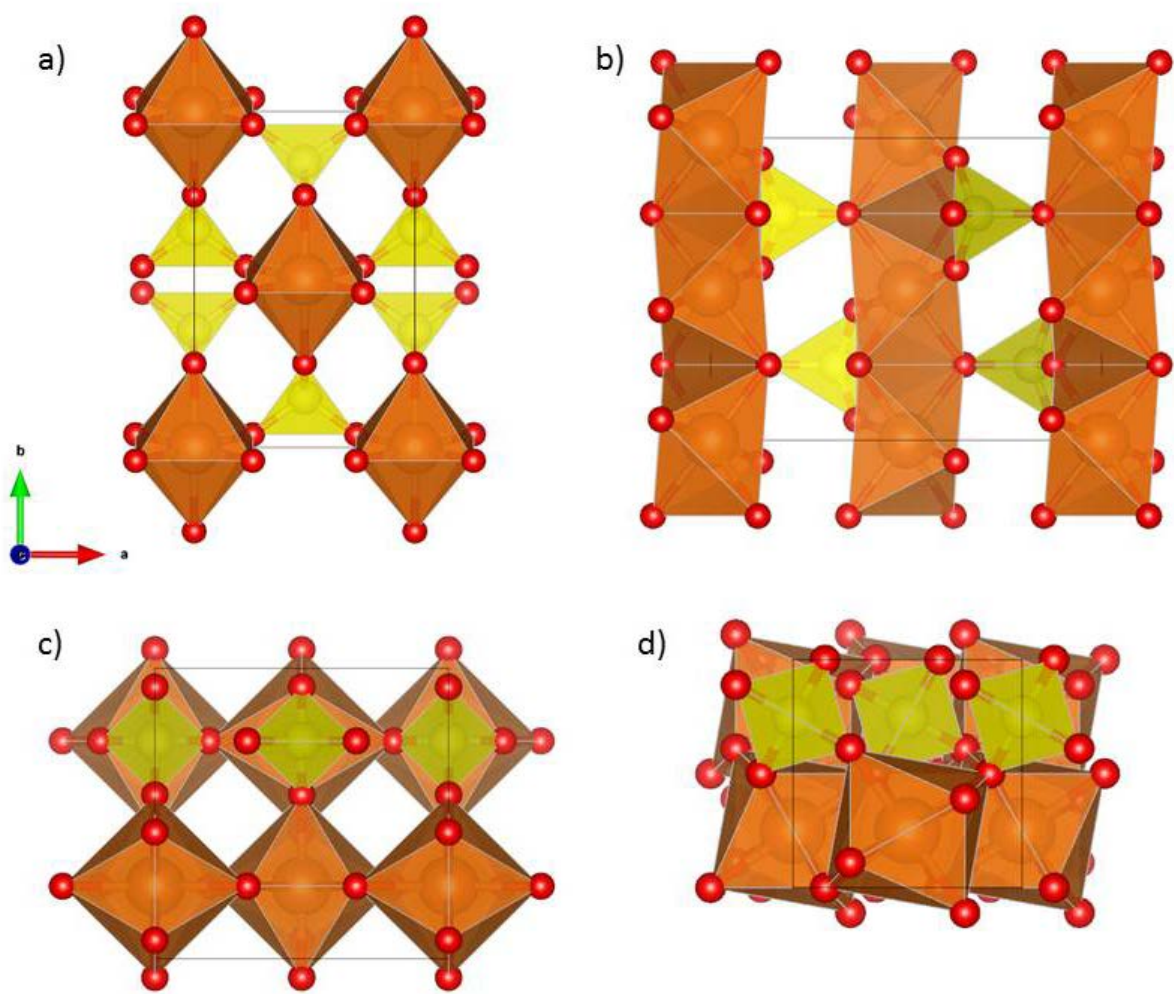

**Fig. 1** Crystal structure of α-MgSO$_4$ (a), β-MgSO$_4$ (b), δ-MgSO$_4$ (c), and ε-MgSO$_4$ (d). The coordination polyhedra of S are shown in yellow and the coordination polyhedra of Mg in orange. The oxygen atoms are shown as small red spheres. Note that δ-MgSO$_4$ and ε-MgSO$_4$ involves an increase of the coordination number of Mg from 6 to 8.

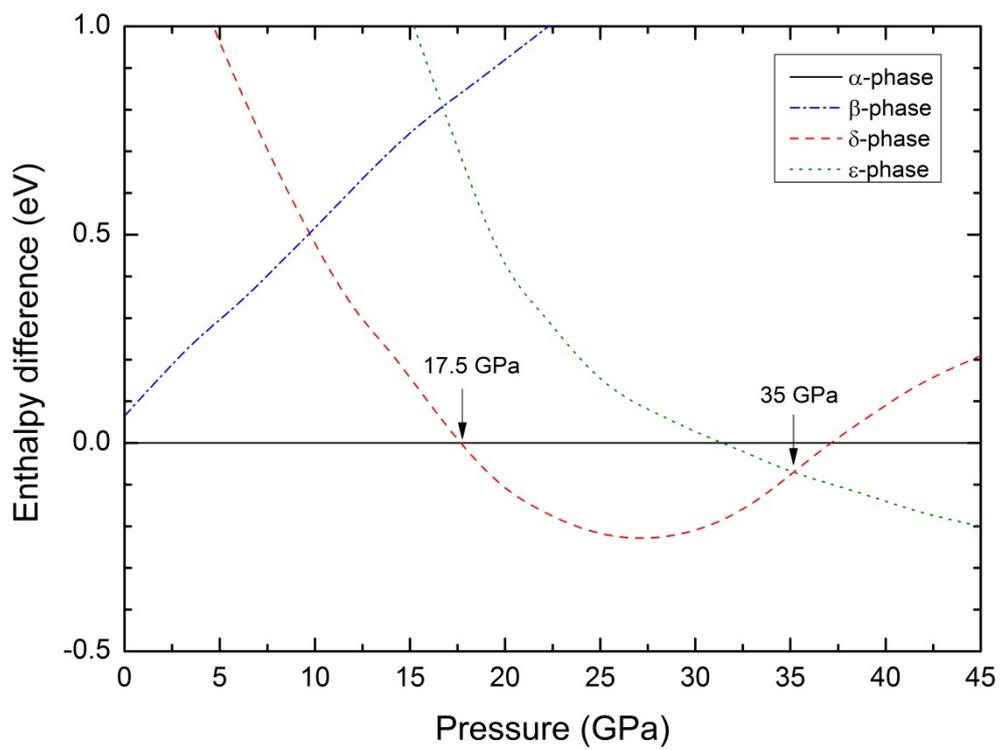

**Fig. 2.** Calculated enthalpy difference versus pressure for the different polymorphs. The enthalpy difference has been calculated relative to α-MgSO$_4$.

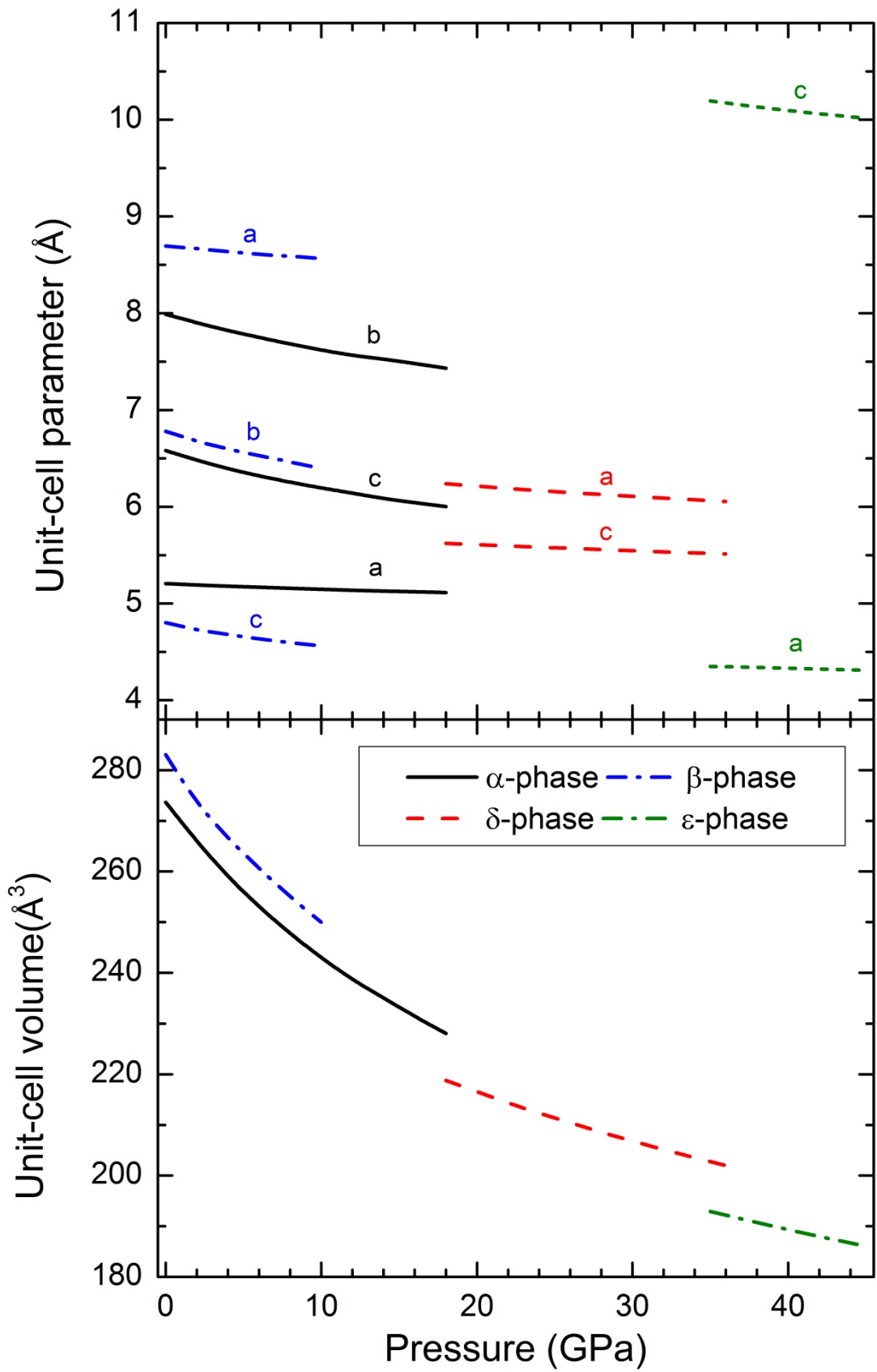

**Fig. 3.** Pressure dependence of the unit-cell parameters and volume for the different polymorphs of MgSO$_4$.

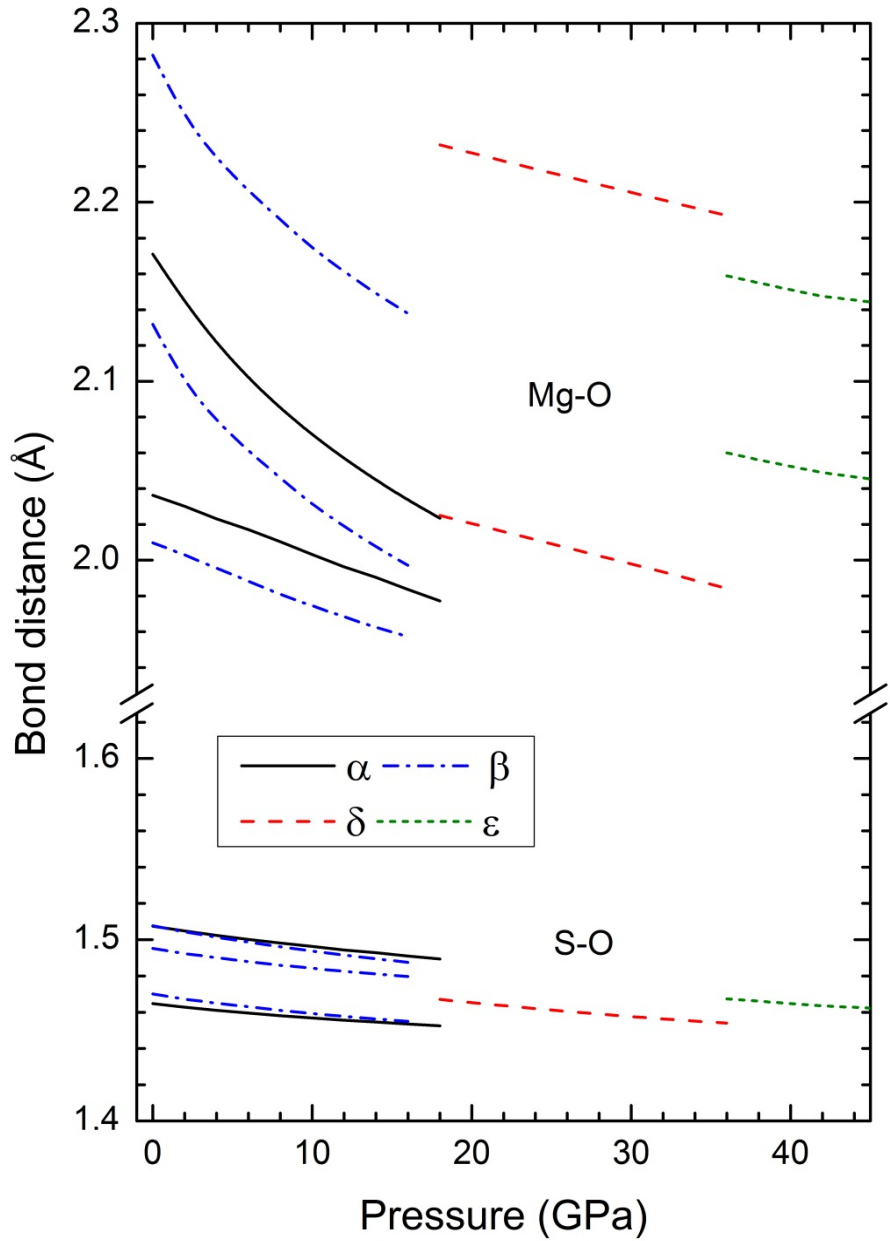

**Fig. 4.** Calculated bond distances versus pressure in different polymorphs of MgSO$_4$

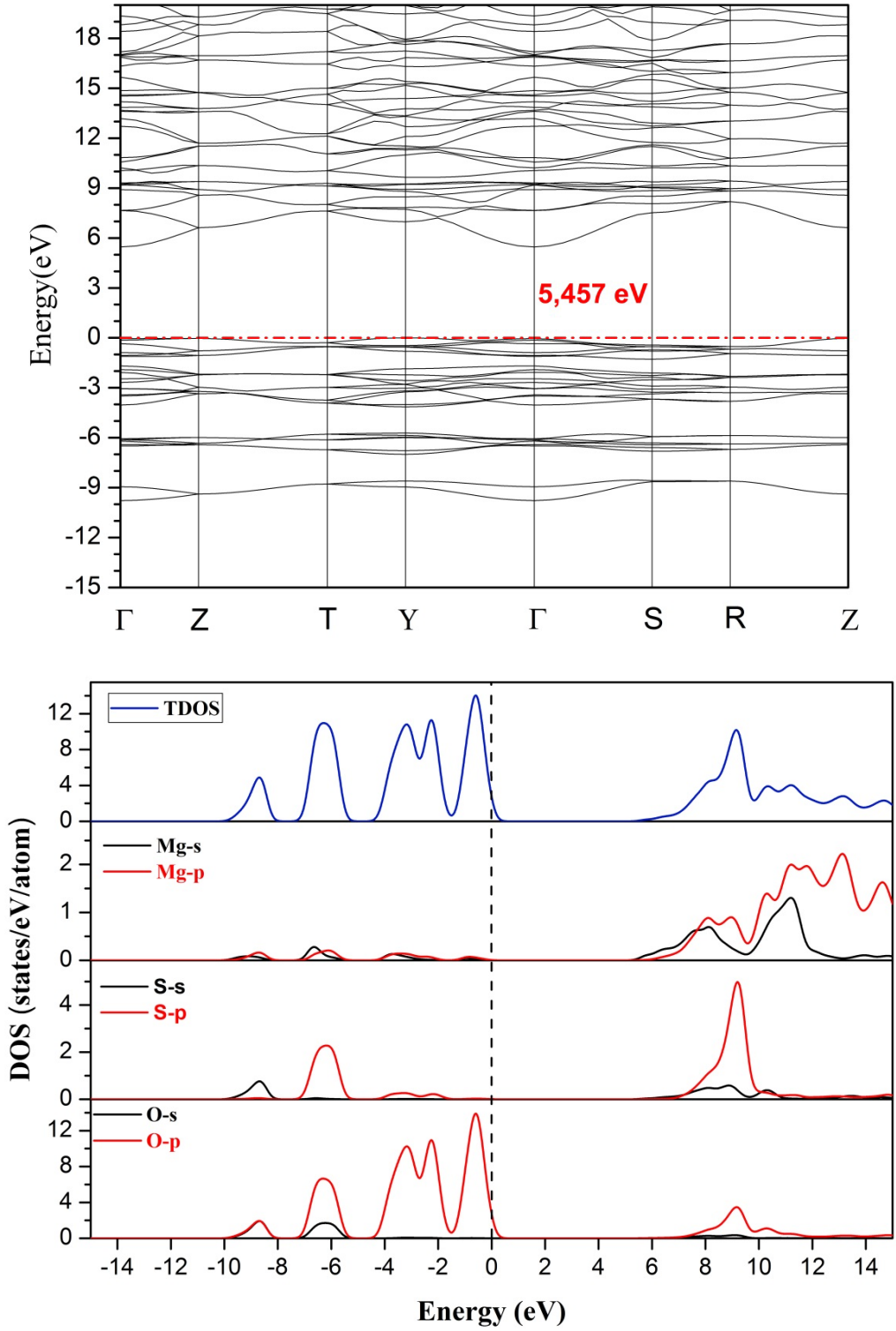

**Fig. 5.** (Top) Calculated band structure of α-MgSO₄ at ambient pressure. (Bottom) Calculated total density of states (TDOS) and partial densities of states for Mg, S, and O.

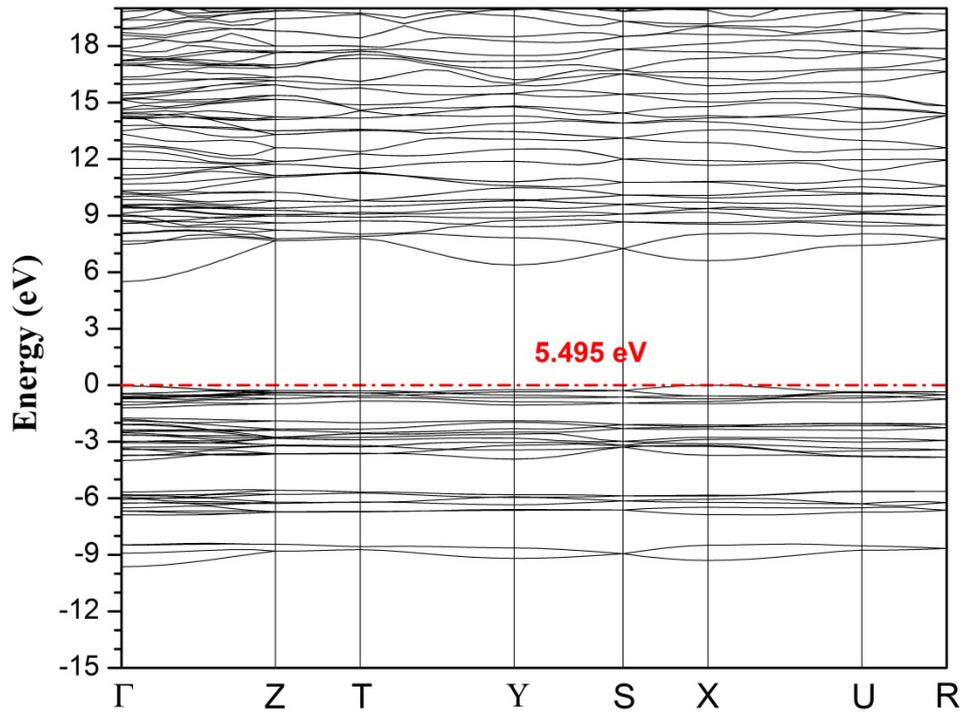

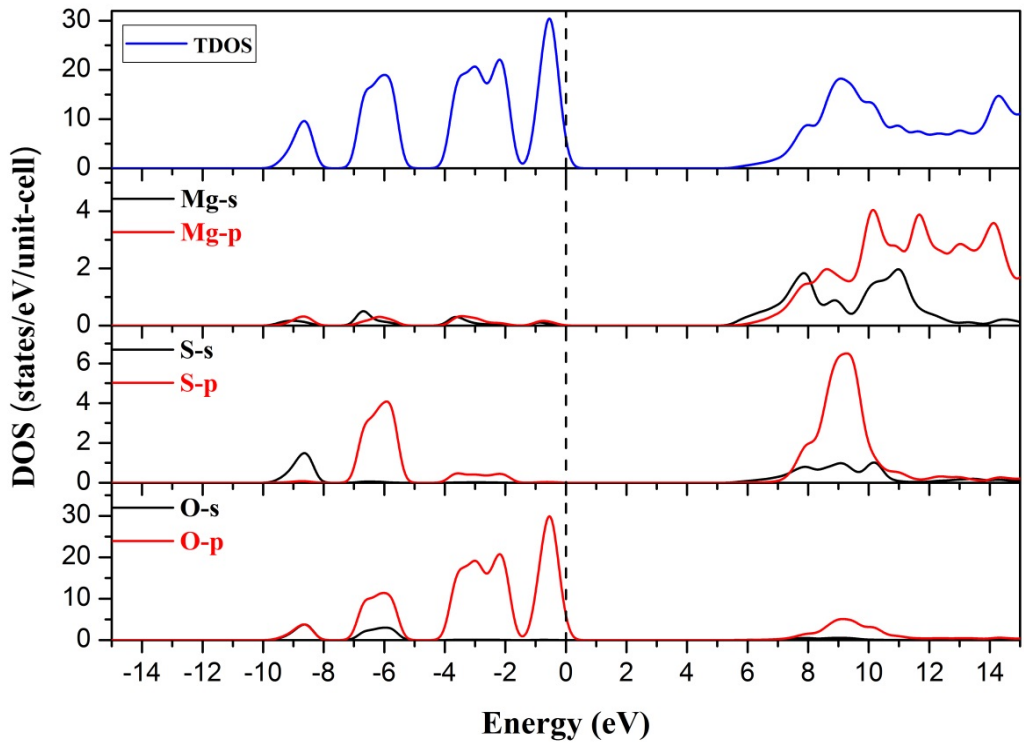

**Fig. 6.** (Top) Calculated band structure of β-MgSO$_4$ at ambient pressure. (Bottom) Calculated total density of states (TDOS) and partial densities of states for Mg, S, and O.

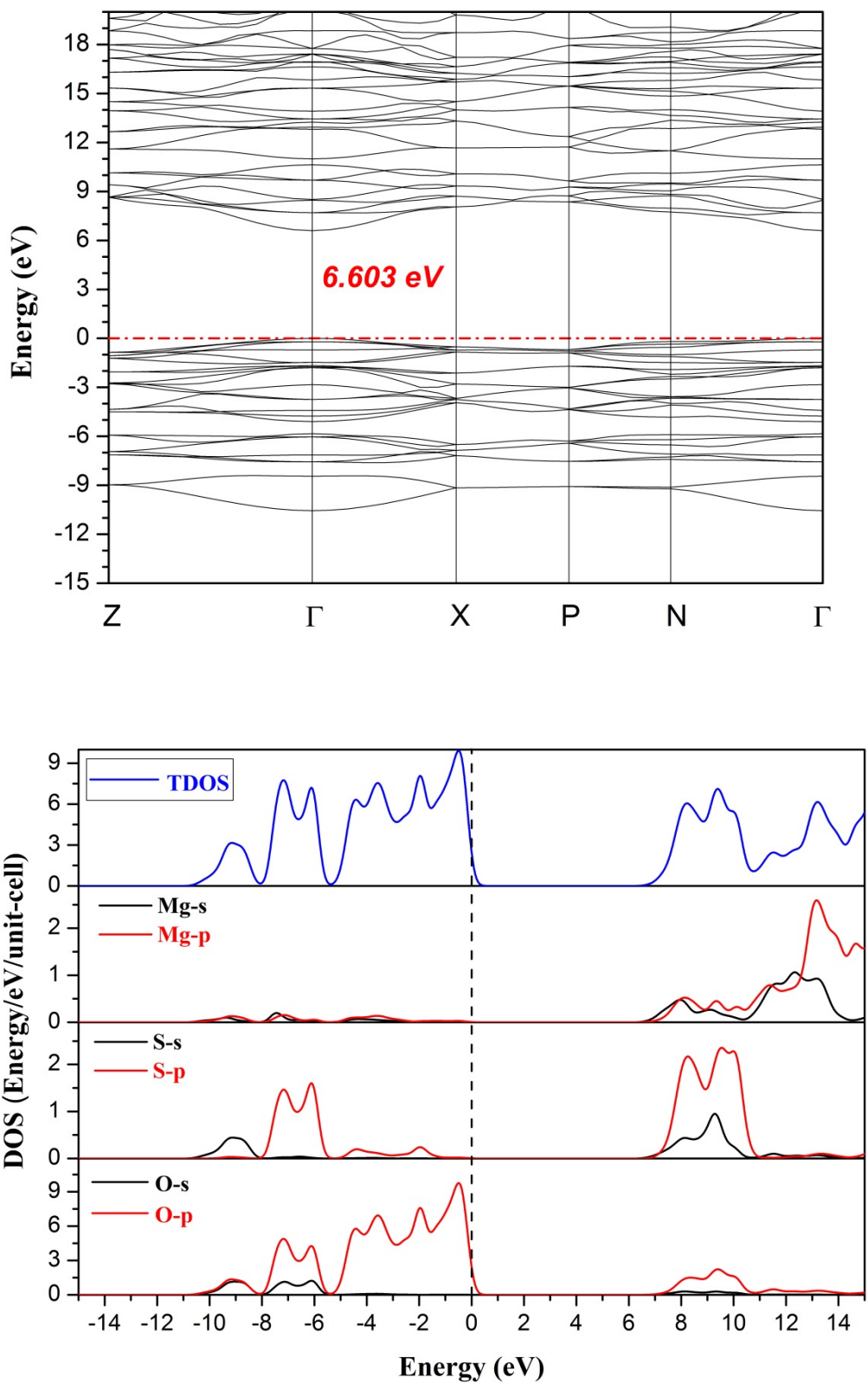

**Fig. 7.** (Top) Calculated band structure of δ-MgSO₄ at ambient pressure. (Bottom) Calculated total density of states (TDOS) and partial densities of states for Mg, S, and O.

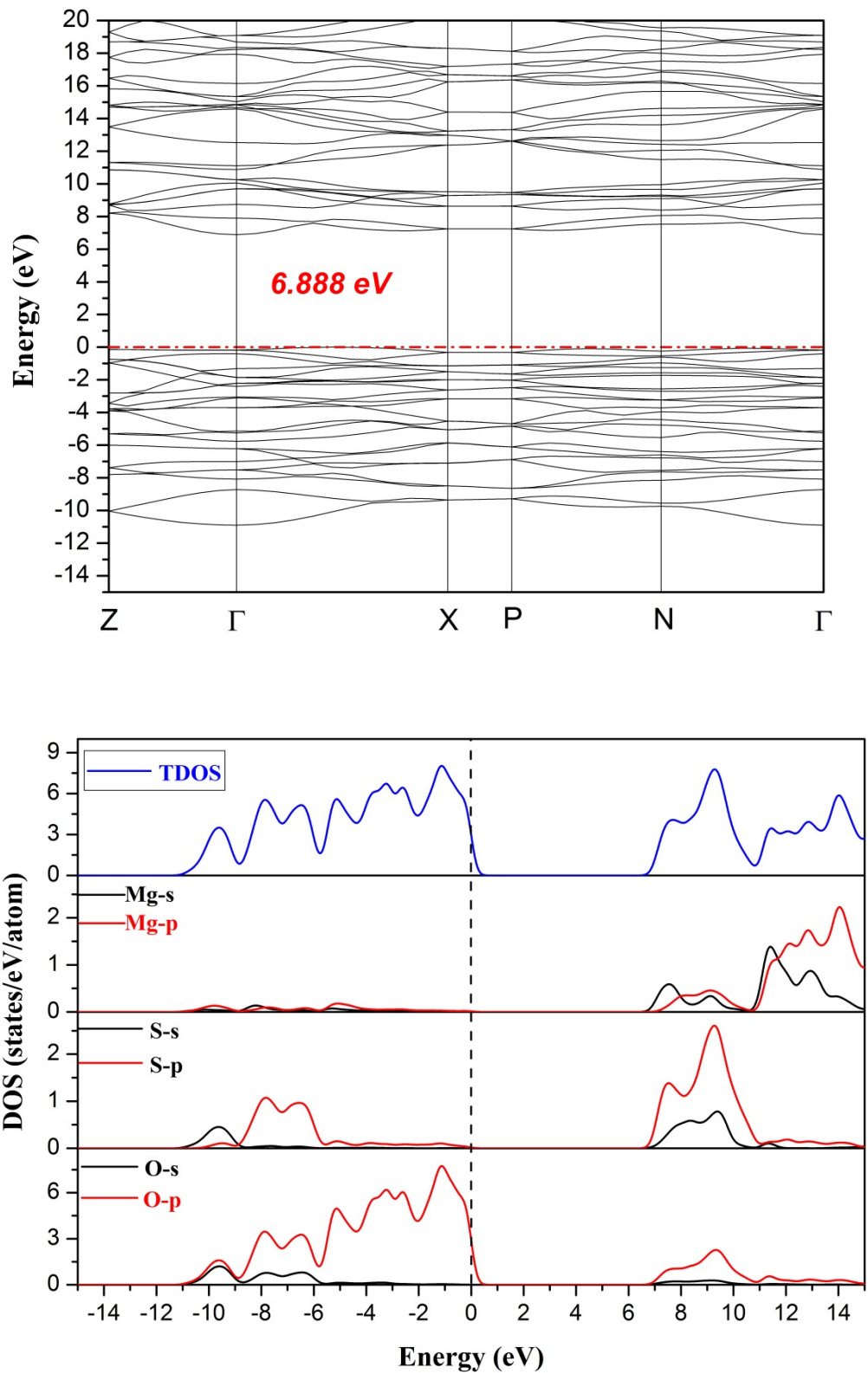

**Fig. 8.** (Top) Calculated band structure of ε-MgSO₄ at ambient pressure. (Bottom) Calculated total density of states (TDOS) and partial densities of states for Mg, S, and O.

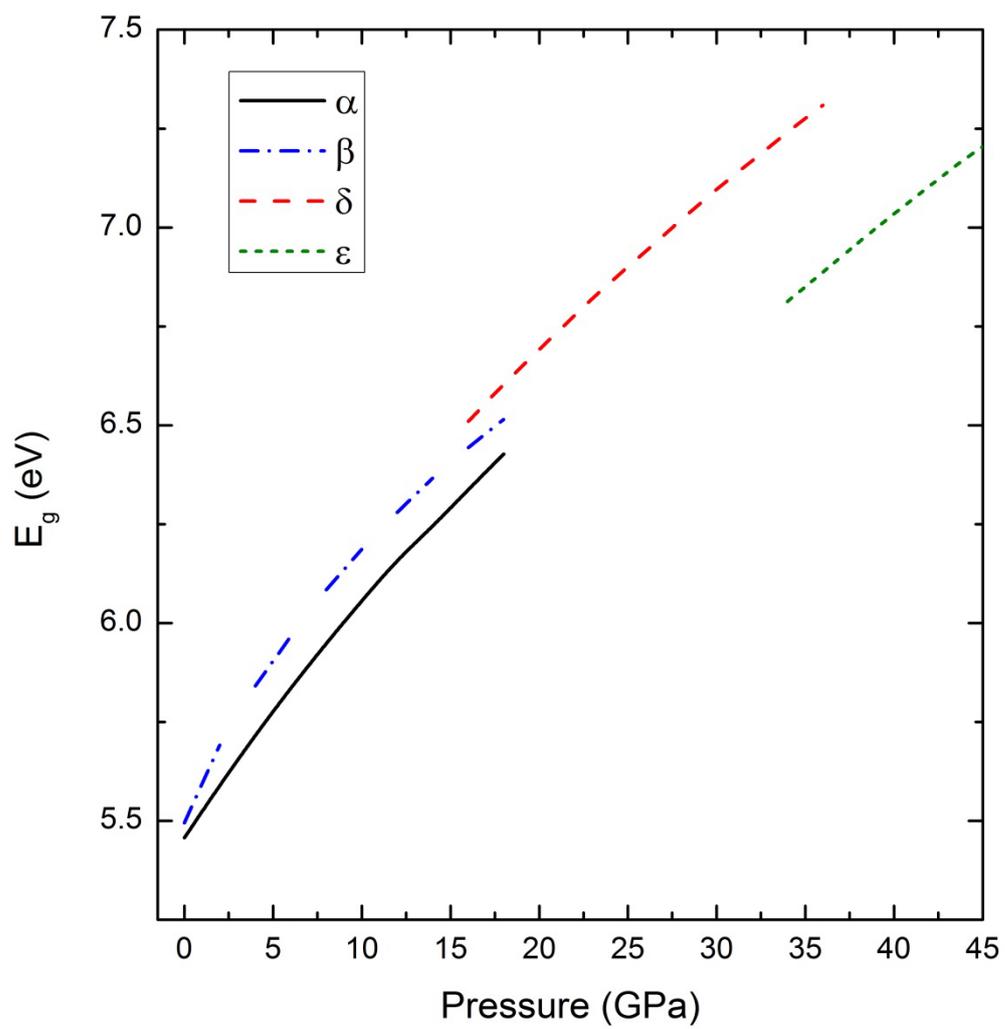

**Fig. 9.** Pressure dependence of the band-gap energy for the different polymorphs of MgSO$_4$.